\documentclass[12pt]{article}
\usepackage[dvips]{graphicx}
\usepackage{latexsym, amssymb}

\newtheorem{Theorem}{Theorem}
\newtheorem{Definition}{Definition}
\newtheorem{Lemma}{Lemma}
\newtheorem{Corollary}[Theorem]{Corollary}

\newtheorem{Example}{Example}
\newtheorem{Remark}{Remark}

\begin{document}
\baselineskip=18pt

\title{Minimum Pseudo-Weight and Minimum Pseudo-Codewords of LDPC Codes
\thanks{This research is supported in part by the National
Natural Science Foundation of China under the Grants 60402031, and
the DSTA research grant R-394-000-025-422. }}

\author{
Shu-Tao Xia\thanks{Shu-Tao Xia is with the Graduate School at
Shenzhen of Tsinghua University, Shenzhen, Guangdong 518055, P. R.
China. E-mail: xiast@sz.tsinghua.edu.cn} and Fang-Wei
Fu\thanks{Fang-Wei Fu is with the Temasek Laboratories, National
University of Singapore, 5 Sports Drive 2, Singapore 117508,
Republic of Singapore (on leave from the Department of
Mathematics, Nankai University, Tianjin 300071, P. R. China).
E-mail: tslfufw@nus.edu.sg}}

\maketitle

\begin{abstract}
In this correspondence, we study the minimum pseudo-weight and
minimum pseudo-codewords of low-density parity-check (LDPC) codes
under linear programming (LP) decoding. First, we show that the
lower bound of Kelley, Sridhara, Xu, and Rosenthal on the
pseudo-weight of a non-zero pseudo-codeword of an LDPC code with
girth greater than 4 is tight if and only if this pseudo-codeword is
a real multiple of a codeword. Then, the lower bound of Kashyap and
Vardy on the stopping distance of an LDPC code is proved to be also
a lower bound on the pseudo-weight of a non-zero pseudo-codeword of
an LDPC code with girth 4, and this lower bound is tight if and only
if this pseudo-codeword is a real multiple of a codeword. Using
these results we further obtain that for some LDPC codes, there are
no other minimum pseudo-codewords except the real multiples of
minimum weight codewords. This means that the LP decoding for these
LDPC codes is asymptotically optimal in the sense that the ratio of
the probabilities of decoding errors of LP decoding and
maximum-likelihood decoding approaches 1 as the signal-to-noise
ratio (SNR) tends to infinity. Finally, some LDPC codes are listed
to illustrate these results.

\end{abstract}

{\bf Index Terms:}\quad LDPC codes, linear programming (LP)
decoding, fundamental cone, pseudo-codewords, pseudo-weight,
stopping sets.

\section{Introduction}

In the study of iterative decoding of low-density parity-check
(LDPC) codes, Wiberg \cite{wib} and Koetter and Vontobel
\cite{koetter} showed that pseudo-codewords play an important role
when characterizing the performance of LDPC codes. Koetter and
Vontobel \cite{koetter} presented an explanation for the relevance
of pseudo-codewords in iterative decoding based on graph covering
and showed that the set of pseudo-codewords can be described by the
so-called fundamental polytope. Recently, linear programming (LP)
decoding of linear codes was introduced by Feldman, Wainwright and
Karger \cite{f}\cite{fwk}. The feasible region of the linear
programming problem in LP decoding \cite{f}\cite{fwk} agrees with
the fundamental polytope. It is known that when characterizing the
performance of linear codes under LP decoding, pseudo-codewords,
especially the pseudo-codewords with minimum pseudo-weight (or
minimum pseudo-codewords for short), also play an important role. In
\cite{dptru}, Di {\sl et al}. showed that the performance of an LDPC
code under message passing decoding over a binary erasure channel is
closely related to the stopping sets in the factor graph. Since the
support of any pseudo-codeword is a stopping set \cite{koetter},
there are some relations between the minimum pseudo-codewords and
the nonempty stopping sets of smallest size
\cite{kv}\cite{sv}\cite{xf}.

Recently, pseudo-codewords and minimum pseudo-weights of binary
linear codes have been studied in \cite{cs}, \cite{f}, \cite{fwk},
\cite{ksxr}-\cite{klvw2}, \cite{skr}, \cite{vk}-\cite{v}, and
\cite{xf}. Chaichanavong and Siegel \cite[Theorem 3]{cs} gave a
lower bound on the pseudo-weight of a non-zero pseudo-codeword of
an LDPC code. Xia and Fu \cite{xf} showed that the
Chaichanavong-Siegel bound is tight if and only if the
pseudo-codeword is a real multiple of a codeword. Using this
result they further obtained that for some LDPC codes, e.g.,
Euclidean plane and projective plane LDPC codes \cite{klf}, there
are no other minimum pseudo-codewords except the real multiples of
minimum weight codewords. Recently, Kelley, Sridhara, Xu, and
Rosenthal \cite[Theorem III.1]{ksxr}\cite[Theorem 3.1]{ks}
presented a lower bound on the pseudo-weight of a non-zero
pseudo-codeword of an LDPC code with girth greater than 4, which
includes the Chaichanavong-Siegel bound as a special case. In
\cite{kv}, Kashyap and Vardy gave a lower bound on the stopping
distance of an LDPC code. In this correspondence, we study the
minimum pseudo-weight and minimum pseudo-codewords of LDPC codes
under LP decoding. The results mentioned in the abstract are
obtained. The rest of this correspondence is organized as follows.
In Section II, we briefly review LP decoding and pseudo-codewords
of binary linear codes. In Section III, the main results of this
correspondence are given and some LDPC codes are listed to
illustrate these results. In Sections IV, the proofs of the main
results are given. In Section V we end with some concluding
remarks.

\section{Preliminaries}

Let $C$ be a binary $[n,k,d]$ linear code with length $n$, dimension
$k$, and minimum distance $d$. The codewords with (Hamming) weight
$d$ are called {\it minimum codewords} of $C$. Let $A_i$ be the
number of codewords of weight $i$. Let $H$ be an $m\times n$
parity-check matrix of $C$, where the rows of $H$ may be dependent.
Let $I=\{1,2,\ldots, n\}$ and $J=\{1,2,\ldots, m\}$ denote the sets
of column indices and row indices of $H$, respectively. The
\emph{Tanner graph} $G_H$ corresponding to $H$ is a bipartite graph
comprising $n$ variable nodes labelled by the elements of $I$, $m$
check nodes labelled by the elements of $J$, and the edge set
$E\subseteq \{(i,j) : i\in I, j\in J\}$, where there is an edge
$(i,j)\in E$ if and only if $h_{ji}=1$. The \emph{girth} $g$ of
$G_H$, or briefly the girth of $H$, is defined as the minimum length
of a cycle in $G_H$. Note that the girth $g$ must be an even integer
not smaller than 4.
\begin{Definition}
A stopping set $S$ is a subset of $I$ such that the restriction of
$H$ to $S$, i.e., the $m\times|S|$ sub-matrix of $H$ consisting of
the columns indexed by $S$, does not contain a row of weight one.
The smallest size of a nonempty stopping set, denoted by $s(H)$, is
called the stopping distance of $C$. A stopping set with size $s(H)$
is called a smallest stopping set. The number of smallest stopping
sets is denoted by $T_s(H)$.
\end{Definition}
In other words, the stopping set $S$ is a subset of variable nodes
in $G_H$ such that all the neighbors of $S$ are connected to $S$ at
least twice. For more results on stopping sets and stopping distance
we refer the readers to \cite{dptru}, \cite{kv}, \cite{ouvz},
\cite{sv}, and \cite{xf2}.

Suppose a codeword $\mathbf{c}$ is transmitted over a binary-input
memoryless channel and $\mathbf{y}$ is the output of the channel.
The log-likelihood ratio vector is defined by
${\lambda}=(\lambda_1, \lambda_2, \ldots, \lambda_n)$ where
$\lambda_i = \ln \frac{\Pr \{y_i \mid c_i=0\}}{\Pr \{y_i \mid
c_i=1\}}$. Let ${\rm conv}(C)$ be the convex hull of $C$ in the
real space ${{\mathbb R}}^n$. Maximum-likelihood (ML) decoding is
equivalent to the following optimization problem
\cite{f}\cite{fwk}: Find $\mathbf{x}\in {\rm conv}(C)$ that
minimizes $\mathbf{\lambda} \mathbf{x}^T$. To decrease the
decoding complexity, the region ${\rm conv}(C)$ should be relaxed.
For each row $\mathbf{h}_j$ of $H$, $1\le j\le m$, let
$C_j=\{\mathbf{c}\in \{0, 1\}^n: \mathbf{h}_j \mathbf{c}^T=0
\;\mbox{mod}\; 2\}.$ The {\it fundamental polytope} of $C$ is
defined as $P(H)=\bigcap_{j=1}^m {\rm conv}(C_j).$ LP decoding
then solves the following optimization problem \cite{f}\cite{fwk}:
Find $\mathbf{x}\in P(H)$ that minimizes $\lambda \mathbf{x}^T$.
Note that ${\rm conv}(C)\subseteq P(H)$. However, usually ${\rm
conv}(C)\subset P(H)$ which implies that the LP decoder is a
sub-optimal decoder. The {\it support} of a real vector
$\mathbf{x}\in {{\mathbb R}}^n$, denoted by
${\rm{supp}}(\mathbf{x})$, is defined as the set of positions of
non-zero coordinates in $\mathbf{x}$, or
${\rm{supp}}(\mathbf{x})=\{i: x_i\ne 0\}$. Assuming that the
channel is a binary-input output-symmetric channel, and given that
the code $C$ is linear, we can without loss of generality assume
that the all-zeros codeword was transmitted. When analyzing the LP
decoder for $C$ it is then sufficient to understand the {\it
fundamental cone} $K(H)$ of $H$ which is the conic hull of the
fundamental polytope $P(H)$. The fundamental cone $K(H)$ can be
characterized as follows \cite{f}\cite{fwk}\cite{koetter}: it is
the set of vectors of $\mathbf{x}=(x_1,\ldots,x_n)\in {{\mathbb
R}}^n$ such that $x_i\ge 0,\;i=1,\ldots,n$ and
\begin{eqnarray}
\label{cone} \forall 1\le
j\le m, \forall \;i\in {\rm{supp}}(\mathbf{h}_j), \sum_{l\in
{\rm{supp}}(\mathbf{h}_j)\setminus \{i\}} x_{l} \ge x_i.
\end{eqnarray}
The elements of $K(H)$ are called {\it pseudo-codewords} of $C$.
Hence, the question of whether the LP decoder succeeds is equivalent
to whether the following optimization problem has the zero vector
$\mathbf{0}$ as its optimal solution: Find $\mathbf{x}\in K(H)$ that
minimizes $\sum_{i=1}^n x_i \lambda_i$. Two pseudo-codewords
$\mathbf{x,y}$ are said to be equivalent if there exists a real
number $\alpha>0$ such that $\mathbf{y}=\alpha \mathbf{x}$. Clearly,
$\mathbf{x}\in K(H)\Leftrightarrow \alpha \mathbf{x}\in K(H)$. For
any $\mathbf{x}\in K(H)$, let $[\mathbf{x}]=\{\alpha \mathbf{x}:
\alpha>0\}$.
\begin{Definition}
A pseudo-codeword $\mathbf{x}$ is said to be internal if there
exists a real number $\beta,\;0<\beta<1$ and
$\mathbf{x}^{(1)},\mathbf{x}^{(2)}\in K(H)\setminus [\mathbf{x}]$
such that $\mathbf{x}=\beta
\mathbf{x}^{(1)}+(1-\beta)\mathbf{x}^{(2)}$. If a non-zero
pseudo-codeword $\mathbf{x}$ is not internal, $[\mathbf{x}]$ is
called an {\it edge} of $K(H)$. Let $M(H)$ denote the set of all
edges of $K(H)$. The pseudo-codewords on edges in $M(H)$ are called
{\it minimal pseudo-codewords}.
\end{Definition}
It is known from \cite{v} and linear programming theory
\cite{cs}\cite{ns} that the behavior of the LP decoder is
completely characterized by $M(H)$ and $|M(H)|$ must be finite for
fixed $C$ and $H$. From now on, we only consider the binary-input
additive white Gaussian noise (AWGN) channel.
\begin{Definition}
The (AWGN) {\it pseudo-weight} of a non-zero real vector
$\mathbf{x}\in {{\mathbb R}}^n$ is defined by $w_P(\mathbf{x})=\|
\mathbf{x}\|_1^2/\| \mathbf{x}\|_2^2$, where $\|
\mathbf{x}\|_1=|x_1|+\cdots+|x_n|$ and
$\|\mathbf{x}\|_2=\sqrt{x_1^2+\cdots+x_n^2}$. Denote by $d_P(H)$
the minimum pseudo-weight of non-zero pseudo-codewords of $C$. The
pseudo-codewords with pseudo-weight $d_P(H)$ are called {\it
minimum pseudo-codewords}. Define the pseudo-weight of an edge
$[\mathbf{x}]\in M(H)$ as the pseudo-weight of $\mathbf{x}$. The
edges with minimum pseudo-weight are called {\it minimum edges}.
The number of minimum edges is denoted by $B_P(H)$.
\end{Definition}
It is not difficult to see from linear programming theory
\cite{ns} that minimum pseudo-codewords are also minimal
pseudo-codewords. Note that the minimal pseudo-codewords in the
same edge have the same pseudo-weight.

Just like $d$ and $A_d$ of a linear code are important for
characterizing the performance of ML decoding, $d_P(H)$ and
$B_P(H)$ are crucial for characterizing the performance of LP
decoding. In order to obtain better performance, we should try to
find a desirable parity-check matrix $H$ to maximize $d_P(H)$ and
then minimize $B_P(H)$. Since the support of every codeword is a
stopping set \cite{sv} and every stopping set supports a
pseudo-codeword \cite{koetter}, it is known that $d_P(H)\le
s(H)\le d$ regardless of the choice of $H$, and $B_P(H)\ge
T_s(H)\ge A_d$ for any $H$ such that $d_P(H)=s(H)=d$. It is well
known that LP decoding is asymptotically optimal, in the sense
that the ratio of the probabilities of decoding errors of LP
decoding and ML decoding approaches 1 as the SNR tends to
infinity, if and only if $d_P(H)=d$ and $B_P(H)= A_d$.

Next, we give an example to illustrate the above concepts.
\begin{Example}
\emph{Let $C$ be a binary $[7,3,4]$ cyclic simplex code. The
parity-check matrix $H$ of $C$ is formed by a $7\times 7$ circulant
matrix, where the first row is $(1\; 1\; 0\; 1\; 0\; 0\; 0)$. $H$
has uniform column weight $3$ and girth 6. The seven non-zero
codewords of $C$ are $(1,0,1,1,1,0,0)$ and its cyclic shifts, each
of which is a minimum codeword. All non-empty stopping sets are
$\{1,3,4,5\}$, $\{2,4,5,6\}$, $\{3,5,6,7\}$, $\{1,4,6,7\}$,
$\{1,2,5,7\}$, $\{1,2,3,6\}$, $\{2,3,4,7\}$, $\{1,2,3,4,5,6,7\}$,
where only the first 7 ones are smallest stopping sets. We choose
one minimal pseudo-codeword as a representative from each edge in
$M(H)$. Then all 14 representatives are $(1,0,1,1,1,0,0)$ and its
cyclic shifts, and $(1,2,1,1,1,2,2)$ and its cyclic shifts, where
only the first 7 ones are minimum pseudo-codewords. Thus, there are
14 edges in $M(H)$ and 7 of which are minimum edges. Clearly, $C$
satisfies $d_P(H)=d=4$ and $B_P(H)= A_d=7$, which implies that LP
decoding is asymptotically optimal for $C$.}
\end{Example}

\section{Main Results}

Let $C$ be a binary $[n,k,d]$ linear code with parity-check matrix
$H$. If the Tanner graph $G_H$ has the girth $g\ge 6$ and $H$ has
uniform column weight $\gamma$, Tanner \cite{tanner} showed that
the minimum distance $d$ fulfills $d\ge d_L$, where
\begin{eqnarray}
\label{dL} d_L = \left\{ \begin{array}{ll}
1+\gamma+\sum_{i=1}^{(g-6)/4}\gamma(\gamma-1)^i, &{g}/{2} \; \rm{odd}, \\
\\
1+\gamma+\sum_{i=1}^{(g-8)/4}\gamma(\gamma-1)^i+(\gamma-1)^{(g-4)/4},&
{g}/{2} \; \rm{even}.
\end{array}
\right.\!\!\!\!\!\!\!\!
\end{eqnarray}
Orlitsky \emph{et al.} \cite{ouvz} further obtained that $d_L$ is
still a lower bound on the stopping distance, i.e., $s(H)\ge d_L$.
Recently, Kelley, Sridhara, Xu, and Rosenthal \cite{ksxr}\cite{ks}
proved that the minimum pseudo-weight satisfies $d_P(H)\ge d_L$, and
the bound still holds when $H$ has non-uniform column weight with
minimum column weight $\gamma$. In the next theorem, which will be
proved in section IV, we give a necessary and sufficient condition
for $w_P(\mathbf{x})=d_L$ to hold for a non-zero pseudo-codeword
$\mathbf{x}\in K(H)$.

\begin{Theorem}
\label{th1} Let $C$ be a binary linear code with length $n$. Let $H$
be a parity-check matrix of $C$ with girth $g\ge 6$ and minimum
column weight $\gamma$. Let $\mathbf{x}=(x_1,\ldots,x_n)\in K(H)$ be
a non-zero pseudo-codeword and $d_L$ be defined in $(\ref{dL})$.
Then $w_{P}(\mathbf{x})= d_L$ if and only if
$d_L\mathbf{x}/\sum_{l=1}^n x_l$ is a codeword of $C$ with weight
$d_L$.
\end{Theorem}

It is easy to check that
\begin{eqnarray}
\label{dL2} d_L = \left\{ \begin{array}{ll}
{\beta}^{(g-2)/4}+2\left(\frac{{\beta}^{(g-2)/4}-1}
{{\beta}-1}\right), &g/2 \; \rm{odd}, \\
2\left(\frac{{\beta}^{g/4}-1}{{\beta}-1}\right),& g/2 \; \rm{even},
\end{array}
\right.
\end{eqnarray}
where ${\beta}=\gamma-1$. In particular,
\begin{eqnarray}
\label{dL3} d_L = \left\{ \begin{array}{ll}
{\beta}+2, &g=6, \\
2({\beta}+1), &g=8, \\
\beta^2+2{\beta}+2, &g=10, \\
2({\beta}^2+{\beta}+1),&g=12, \\
\beta^3+2\beta^2+2{\beta}+2, &g=14, \\
2({\beta}^3+{\beta}^2+{\beta}+1),\quad\quad&g=16.
\end{array}
\right.
\end{eqnarray}
Clearly, for the codes satisfying the conditions of Theorem
$\ref{th1}$, $d=d_L$ will imply $d_P(H)=s(H)=d$. Furthermore, by
Theorem \ref{th1}, we have the next corollary.

\begin{Corollary}
\label{co1} Let $C$ be a binary $[n,k,d]$ linear code with length
$n$, dimension $k$ and minimum distance $d$. Let $H$ be a
parity-check matrix of $C$ with girth $g$ ($g\geq 6$) and minimum
column weight $\gamma$. If $d=d_L$, where $d_L$ is defined in {\rm
(\ref{dL})}, then $B_P(H)=T_s(H)=A_d$, where $A_d$ is the number
of minimum codewords, $T_s(H)$ is the number of smallest stopping
sets, and $B_P(H)$ is the number of minimum edges.
\end{Corollary}

\begin{Remark}
For a code $C$ satisfying the conditions of Corollary $\ref{co1}$,
the minimum codewords, the nonempty stopping sets of smallest size
and the minimum edges are all equivalent, which implies that LP
decoding is asymptotically optimal for $C$.
\end{Remark}

In Example 1, $\gamma=3$, $g=6$, $d=4$, and $d_L=1+\gamma=4=d$.
Hence, $C$ satisfies the conditions of Theorem \ref{th1} and
Corollary \ref{co1}, and $B_P(H)=T_s(H)=A_d=7$.

Note that {\rm \cite[Theorems 1 and 2]{xf}} are the special case of
$g=6$ of Theorem $\ref{th1}$ and Corollary $\ref{co1}$,
respectively. In \cite{xf}, it is shown that two classes of finite
geometry LDPC codes, i.e., the projective plane LDPC codes and
Euclidean plane LDPC codes \cite{klf}, meet the conditions of
Corollary \ref{co1}. Thus, LP decoding is asymptotically optimal for
finite plane LDPC codes. Below we give some more examples of LDPC
codes satisfying the conditions of Corollary $\ref{co1}$.

\begin{Example}
\emph{A class of regular LDPC codes called $LU(3,q)$ codes were
constructed in \cite{kpppf}, where $q$ is a prime power. $LU(3,q)$
codes have the following parameters, where $n$ is the code length,
$d$ is the minimum distance, $m$ is the number of rows of
the parity-check matrix, $\rho$ is the uniform row weight of the
parity-check matrix, $\gamma$ is the uniform column weight of the
parity-check matrix, and $g$ is the girth of the Tanner graph.}
$$n=q^3, \;m=q^3, \;d= 2q, \;\rho=q, \;\gamma=q, \;g=8.$$
\emph{This class of LDPC codes meet the conditions of Corollary
\ref{co1}. Thus, LP decoding is asymptotically optimal for
$LU(3,q)$ codes.}
\end{Example}

\begin{Example}
\emph{ In \cite{lp} \cite{vt}, regular LDPC codes were constructed
from generalized polygons. In \cite[Table 1]{lp}, for a prime
power $q$, LDPC codes $W(q)$, $H(3,q^2)$, $H(q)$, $T(q^3,q)$, and
$\bar O(q)$ have the following parameters, where $n$ is the code
length, $d$ is the minimum distance, $\gamma$ is the uniform
column weight of the parity-check matrix, and $g$ is the girth of
the Tanner graph.}
\begin{description}
\item{{\rm (i)}} $W(q):$ $\;n=(q+1)(q^2+1)$, $d=2(q+1)$, $\gamma=q+1$, $g=8$;
\item{{\rm (ii)}} $H(3,q^2):$ $\;n=(q^2+1)(q^3+1)$, $d=2(q+1)$, $\gamma=q+1$, $g=8$;
\item{{\rm (iii)}} $H(q):$ $\;n=(q+1)(q^4+q^2+1)$, $d=2(q^2+q+1)$, $\gamma=q+1$,
$g=12$;
\item{{\rm (iv)}} $T(q^3,q):$ $\;n=(q^3+1)(q^8+q^4+1)$, $d=2(q^2+q+1)$,
$\gamma=q+1$, $g=12$;
\item{{\rm (v)}} $\bar O(q), q=2^{2e+1}:$ $\;n=(q^2+1)(q^3+1)(q^6+1)$,
$d=2(q^3+q^2+q+1)$, $\gamma=q+1$, $g=16$.
\end{description}
\emph{By (\ref{dL3}), it is obvious that these LDPC codes meet the
conditions of Corollary \ref{co1}. Thus, LP decoding is
asymptotically optimal for them. }
\end{Example}

\begin{Example}
\emph{In \cite{skr}, some LDPC codes with $d_P(H)=d$ were
constructed by enumerating a regular tree for a fixed number $l$ of
layers and employing a connection algorithm based on mutually
orthogonal Latin
squares to close the tree. \\
(i) \emph{Type-I A} construction \cite{skr}: It is known that if
$l$ or $g/2$ is odd, then $d=d_L$. Hence, these LDPC codes with
odd $g/2$ meet the conditions of Corollary \ref{co1} and LP
decoding is asymptotically optimal for them.\\
(ii) \emph{Type-II} construction \cite{skr}: For the binary case
and $l=3$, the Type II construction yields exactly the projective
plane LDPC codes \cite{txla05}\cite{xf}. For the binary case and
$l=4$, it is conjectured that $d=d_L$ in \cite{skr}. Clearly, if
this conjecture is true, then these LDPC codes meet the conditions
of Corollary \ref{co1} and LP decoding is asymptotically optimal
for them. In particular, it is known from \cite{skr} and \cite{vt}
that this is true for the
$(2,2)$-Finite-Generalized-Quadrangles-based LDPC codes.}
\end{Example}

The next theorem shows that the lower bound of Kashyap and Vardy
\cite{kv} on the stopping distance of an LDPC code is also a lower
bound on the pseudo-weight of a non-zero pseudo-codeword of this
LDPC code, and this lower bound is tight if and only if this
pseudo-codeword is a real multiple of a codeword. The proof of this
theorem will be given in section IV.

\begin{Theorem}
\label{th3} Let $C$ be a binary linear code with length $n$. Let
$H$ be an $m\times n$ parity-check matrix of $C$ with minimum
column weight $\gamma$. If any two distinct columns of $H$ have at
most $\lambda$ common $1$'s and ${\gamma}/{\lambda}$ is an
integer, then $w_{P}(\mathbf{x})\ge \frac{\gamma}{\lambda}+1$ for
any non-zero pseudo-codeword $\mathbf{x}=(x_1,\ldots,x_n)\in
K(H)$. Moreover, equality holds if and only if
$(\frac{\gamma}{\lambda}+1)\mathbf{x}/\sum_{l=1}^n x_l$ is a
codeword of $C$ with weight $\frac{\gamma}{\lambda}+1$.
\end{Theorem}

\begin{Remark}
If $H$ has uniform column weight $\gamma$, Kashyap and Vardy {\rm
\cite[Theorem 1]{kv}} showed that the stopping distance $s(H) \geq
\frac{\gamma}{\lambda}+1$. Since $d_P(H)\le s(H)$, Theorem {\rm
\ref{th3}} implies the Kashyap-Vardy lower bound on the stopping
distance.
\end{Remark}

Clearly, for codes satisfying the conditions of Theorem $\ref{th3}$,
$d=\frac{\gamma}{\lambda}+1$ will imply $d_P(H)=s(H)=d$.
Furthermore, by Theorem \ref{th3}, we obtain the next corollary.
\begin{Corollary}
\label{co2} Let $C$ be a binary $[n,k,d]$ linear code with length
$n$, dimension $k$ and minimum distance $d$. Let $H$ be a
parity-check matrix of $C$ with minimum column weight $\gamma$. If
any two distinct columns of $H$ have at most $\lambda$ common
$1$'s and ${\gamma}/{\lambda}$ is an integer, and if
$d=\frac{\gamma}{\lambda}+1$, then $B_P(H)=T_s(H)=A_d$, where
$A_d$ is the number of minimum codewords, $T_s(H)$ is the number
of smallest stopping sets, and $B_P(H)$ is the number of minimum
edges.
\end{Corollary}

\begin{Remark}
In Theorem $\ref{th3}$ and Corollary $\ref{co2}$, the girth of the
Tanner graph $G_H$ is at least $6$ if $\lambda=1$ and $4$ if
$\lambda>1$. The special case of $\lambda=1$ in Theorem $\ref{th3}$
and Corollary $\ref{co2}$ are exactly {\rm\cite[Theorems 1 and
2]{xf}} and the special case of $g=6$ in Theorem $\ref{th1}$ and
Corollary $\ref{co1}$.
\end{Remark}

\begin{Example}
\emph{ Consider the binary $[2^r-1,2^r-r-1,3]$ Hamming code. Let $H$
be a parity-check matrix which consists of all the non-zero
codewords of the binary $[2^r-1,r,2^{r-1}]$ simplex code. It is easy
to see that $\gamma=2^{r-1}$ and $\lambda=2^{r-2}$. Hence, by
Theorem \ref{th3} and Corollary \ref{co2}, $d_P(H)=3$ and any
minimum pseudo-codeword must be the real multiple of some minimum
codeword. }
\end{Example}

Let $q=2^s$ and $EG(m,q)$ be the $m$-dimensional Euclidean Geometry over $GF(q)$.
It is known from \cite{lc} and \cite{txla05} that there are $q^m$
points and $q(q^m-1)/(q-1)$ hyperplanes in $EG(m,q)$. By removing
a point of $EG(m,q)$ together with the $(q^m-1)/(q-1)$ hyperplanes
containing this point, we obtain a slightly modified incidence
matrix $H$ of points and hyperplanes in $EG(m,q)$. Suppose the
rows of $H$ indicate the hyperplanes. The point-hyperplane
Euclidean geometry LDPC code $C$ with the parity-check matrix $H$
has the following parameters: length $n=q^m-1$, uniform column
weight of $H$ $\gamma=q^{m-1}$, uniform row weight $\rho=q^{m-1}$,
girth of Tanner graph $g=4$ if $m>2$. It is easy to see that any
two distinct columns of $H$ have at most $\lambda=q^{m-2}$ common
$1$'s. By Theorem \ref{th3}, we have that $d\ge s(H)\ge d_P\ge
\gamma/\lambda+1= q+1$. From \cite{lc} and \cite{txla05}, we know
that $H$ can be put in cyclic form and the generator polynomial
$g(x)$ can be determined. By \cite[p. 315, (8.33)]{lc}, it is
known that the dimension $k=2^{sm}-(m+1)^s$. The following
examples show that $d=q+1$ in some cases.

\begin{Example}
\emph{ Let $m=3$ and $s=2$. Then $C$ is a binary $[63, 48]$ code
with generator polynomial
$g(x)=1+x^2+x^4+x^{11}+x^{13}+x^{14}+x^{15}$ \cite[pp.
310-311]{lc}. By Theorem \ref{th3}, we know that $d\ge s(H)\ge d_P
\ge q+1=5$. In fact, it is easy to calculate by computer that $d$
does equal 5. For example, $1+x^{23}+x^{33}+x^{36}+x^{37}$ is a
weight-5 codeword. Hence, by Corollary \ref{co2}, we have that
$A_d=T_s(H)=B_P(H)$.  }
\end{Example}

\begin{Example}
\emph{ Let $m=3$ and $s=3$. Then $C$ is a binary $[511,448]$ code
\cite[Example 1]{txla05}. The girth of the Tanner graph is $4$ and
$C$ performs very well under iterative decoding \cite{txla05}. By
Theorem \ref{th3}, we know that $d\ge s(H)\ge d_P \ge q+1=9$. In
fact, it can be calculated by the method in \cite{hfe} that $d=9$.
Hence, by Corollary \ref{co2}, we have that $A_d=T_s(H)=B_P(H)$
and LP decoding for $C$ is asymptotically optimal. }
\end{Example}

\section{ Proofs of Theorems \ref{th1} and \ref{th3}}

In this section, we prove Theorems \ref{th1} and \ref{th3}.
Chaichanavong and Siegel \cite[Proposition 2]{cs} gave a lower bound
on the pseudo-weight of a real vector. In \cite{xf}, the necessary
and sufficient condition for this bound being tight is discussed.
Let $u$ be a positive integer. Denote ${\cal F}_u$ the set of
vectors $\mathbf{y}\in [0, 1/u]^n$ such that $\sum_{i=1}^n y_i=1$.
\begin{Lemma}
\label{lem1} \rm{\cite{xf}} For any $\mathbf{y}\in {\cal F}_u$, we
have $w_P(\mathbf{y})\ge u$. Equality holds if and only if
$\mathbf{y}$ has exactly $u$ non-zero components with value $1/u$.
\end{Lemma}

\subsection*{A. Proof of Theorem \ref{th1}}

From the proof of \cite[Theorem 3.1]{ks}, we know that for any
non-zero $\mathbf{x}=(x_1,\ldots,x_n)\in K(H)$, $d_L x_i\le
\sum_{j=1}^n x_j$ for any $i\in {\rm{supp}}(\mathbf{x})$. Let
$\mathbf{y}=\mathbf{x}/\sum_{i=1}^n x_i$. Then $\mathbf{y}\in
K(H)$ and $\mathbf{y}\in {\cal F}_{d_L}$. Hence, by Lemma
\ref{lem1}, $w_P(\mathbf{x})=w_P(\mathbf{y})\ge d_L$, where
equality holds if and only if
$\mathbf{c}=d_L\mathbf{x}/\sum_{l=1}^n x_l$ is a binary vector
with weight $d_L$. Now, we show that $w_P(\mathbf{x})=d_L$ if and
only if $\mathbf{c}\in C$ and $w_H(\mathbf{c})=d_L$. If
$\mathbf{c}\in C$ and $w_H(\mathbf{c})=d_L$, then
$w_P(\mathbf{x})=w_P(\mathbf{c})=w_H(\mathbf{c})=d_L$. On the
other hand, if $w_P(\mathbf{x})=d_L$, then the pseudo-codeword
$\mathbf{c}$ is a binary vector with weight $d_L$. Next, we show
that $\mathbf{c}\in C$.

\begin{figure}
\begin{center}
\includegraphics[width=2.7 in]{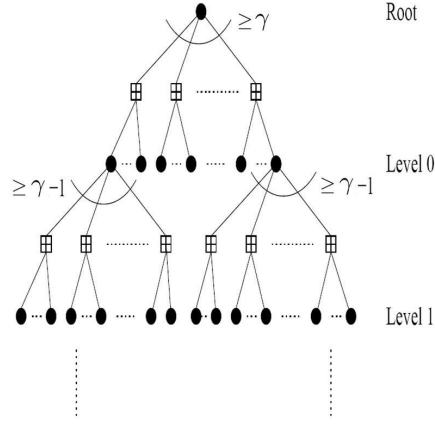}
\label{fig:Pw} \caption{Local tree structure for a
$\ge\gamma$-left graph}
\end{center}
\end{figure}

Clearly, $S={\rm{supp}}(\mathbf{c})$ is a stopping set with size
$d_L$ since $\mathbf{c}$ is a pseudo-codeword. For any fixed $i\in
S$, we construct a local tree of $i$ (see Figure 1) as in the proof
of \cite[Theorem 3.1]{ks}. For the sake of convenience, we briefly
describe the construction procedure as follows. Below we use $f, e$
to denote check nodes and $i,j$ to denote variable nodes of the
Tanner graph $G_H$. Let $t=\lfloor(g-6)/4\rfloor\ge 0$, where
$\lfloor x\rfloor$ is the floor function which denotes the maximum
integer not greater than $x$. Then $g=4t+6$ for odd $g/2$ and
$g=4t+8$ for even $g/2$. In the local tree of $i$, $i$ is the root
of the tree. A check node $f$ connected to $i$ is called a child of
$i$, and a variable node $j$ connected to $f$ except its parent $i$
is called a child of $f$ or a grandchild of $i$, and a check node
$e$ connected to $j$ except its parent $f$ is called a child of $j$,
and so on. For a variable node $j$ in the local tree, let ${\rm
child}(j)$ and ${\rm grch}(j)$ denote the sets of all children and
grandchildren of $j$ respectively. Note that
\begin{eqnarray}
\label{grch} {\rm grch}(j)&=&\bigcup_{f\in {\rm child}(j)} {\rm
child}(f).
\end{eqnarray}
All nodes in $L_0(i)= {\rm grch}(i)$ are called \emph{Level}-0
variable nodes. For $m=1,2,\ldots,t$, all nodes in
\begin{eqnarray}
\label{lm} L_m(i)= \bigcup_{j\in L_{m-1}(i)}{\rm grch}(j)
\end{eqnarray}
are called \emph{Level}-$m$ variable nodes. Fixing a check node
$f^*\in {\rm child}(i)$, denote $N_0(f^*)={\rm child}(f^*)$ and
\begin{eqnarray}
\label{nm} N_{m}(f^*)&=&\bigcup_{j\in N_{m-1}(f^*)}{\rm grch}(j),
\;\;\; m=1,2,\ldots, t+1.
\end{eqnarray}
The local tree of $i$ has $t$ levels if $g=4t+6$ and $t+1$ levels
if $g=4t+8$, where $N_{t+1}(f^*)$ is the set of $(t+1)$-th level
nodes. Since the Tanner graph $G_H$ has girth $g\ge 6$, the local
tree of $i$ has the following pairwise disjoint properties: (i)
all ${\rm child}(f)$ in the union of (\ref{grch}) are pairwise
disjoint, and all ${\rm grch}(f)$ in the union of (\ref{lm}) are
pairwise disjoint; (ii) if $g=4t+6$, $\{i\}, L_0(i),\ldots,
L_t(i)$ are pairwise disjoint; (iii) if $g=4t+8$, all ${\rm
grch}(f)$ in the union of (\ref{nm}) are pairwise disjoint, and
$\{i\}, L_0(i),\ldots,L_t(i)$, $N_{t+1}(f^*)$ are pairwise
disjoint.

Since there are at least $\gamma$ $1$'s in every column of $H$,
from the construction we have that
\begin{eqnarray}
\label{c0} |{\rm child}(i)|\ge \gamma \quad\mbox{and}\quad |{\rm
child}(j)|\ge \gamma-1
\end{eqnarray}
for any intermediate variable node $j$ in the local tree of $i$. Let
$j$ be a variable node which has some children in the local tree of
$i$. Suppose $j\in S$. For each check node $f\in {\rm child}(j)$,
$|{\rm child}(f)\cap S| \ge 1$ since $S$ is a stopping set including
$j$. Thus, noting that $i\in S$, by (\ref{c0}) and the pairwise
disjoint properties, we have
\begin{eqnarray}
\label{S1} |L_0(i)\cap S|=|{\rm grch}(i)\cap S|\ge \gamma;\quad
|{\rm grch}(j)\cap S|\ge \gamma-1 \;\mbox{ if }\; j\in
S\setminus\{i\}.
\end{eqnarray}
Moreover, a necessary condition for equality in $|L_0(i)\cap
S|=\gamma$ is that for each $f\in {\rm child}(i)$, $|{\rm
child}(f)\cap S| = 1$. In other words, for any row $\mathbf{h}$ of
$H$ whose $i$-th component is 1, $w_H(\mathbf{h}_S)=2$ where
$\mathbf{h}_S$ is the restriction of $\mathbf{h}$ to $S$.
Furthermore, by (\ref{c0}), (\ref{S1}) and the pairwise disjoint
properties, for $m=1,2, \ldots, t$,
\begin{eqnarray*}
|L_m(i)\cap S|&=& \bigg|\bigcup_{j\in L_{m-1}(i)}{\rm grch}(j)\cap
S\bigg| \ge \bigg|\bigcup_{j\in
L_{m-1}(i)\cap S}{\rm grch}(j)\cap S\bigg| \\
&\ge& (\gamma-1)|L_{m-1}(i)\cap S|\ge \cdots\ge
(\gamma-1)^m|L_{0}(i)\cap S|\ge (\gamma-1)^m\gamma
\end{eqnarray*}
and if $g=4t+8$,
\begin{eqnarray*}
|N_{t+1}(f^*)\cap S| &=& \bigg|\bigcup_{j\in N_{t}(f^*)}{\rm
grch}(j)\cap S\bigg|
\ge (\gamma-1)|N_{t}(f^*)\cap S| \\
&\ge& \cdots\ge (\gamma-1)^{t+1}|{\rm child}(f^*)\cap S|\ge
(\gamma-1)^{t+1}.
\end{eqnarray*}
In other words, $|\{i\}\cap S| = 1$, $|L_0(i)\cap S| \ge \gamma$,
$|L_1(i)\cap S| \ge \gamma(\gamma-1)$, $\ldots$, $|L_t(i)\cap S| \ge
\gamma(\gamma-1)^t$, and $|N_{t+1}(f^*)\cap S| \ge (\gamma-1)^{t+1}$
if $g=4t+8$. Therefore, we have $|S|\ge d_L$ by adding the above
inequalities and using the pairwise disjoint properties, where a
necessary condition of $|S|=d_L$ is that $|L_0(i)\cap S|=\gamma$,
that is, $w_H(\mathbf{h}_S)=2$ for any row $\mathbf{h}$ of $H$ whose
$i$-th component is 1. This implies that $\mathbf{c}$ satisfies all
the parity-check equations corresponding to the rows $\mathbf{h}$ of
$H$ whose $i$-th component is 1. Thus, when $i$ varies in
$S={\rm{supp}}(\mathbf{c})$, $\mathbf{c}$ must satisfy every
parity-check equation in $H$, i.e., $\mathbf{c}$ is a codeword.

\subsection*{B. Proof of Theorem \ref{th3}}

Let $\mathbf{y}=\mathbf{x}/\sum_{j=1}^n x_j$. Since $\mathbf{x}\in
K(H)$, then $\mathbf{y}\in K(H)$. For fixed $j$, $1\le j\le n$,
let $\mathbf{h}_{q_1}, \mathbf{h}_{q_2},\ldots,
\mathbf{h}_{q_{\gamma}}$ be the rows of $H$ whose $j$-th
components are $1$, i.e., $\mathbf{h}_{q_i}=(h_{q_i,1}, \ldots,
h_{q_i,n})$ and $h_{q_i,j}=1$ for $1\le i\le \gamma$. Since
$\mathbf{y}\in K(H)$, $y_j\le \sum_{l\ne j} y_l h_{q_i, l}, \quad
i=1,\ldots,\gamma$. Hence,
\begin{eqnarray*}
\gamma y_j \le \sum_{i=1}^{\gamma} \sum_{l\ne j} y_l h_{q_i, l} =
\sum_{l\ne j} y_l \left (\sum_{i=1}^{\gamma} h_{q_i, l} \right ).
\end{eqnarray*}
For any $l\ne j$, since the $l$-th column and $j$-th column of $H$
have at most $\lambda$ common $1$'s and $h_{q_i,j}=1$ for $1\le
i\le \gamma$, we have that $\sum_{i=1}^{\gamma} h_{q_i, l} \le
\lambda$. Therefore,
\begin{eqnarray*}
\gamma y_j \le \lambda \sum_{l\ne j} y_l = \lambda (1-y_j), \quad
\mbox{i.e.,}\quad y_j \le \frac{\lambda}{\gamma+\lambda},
\end{eqnarray*}
which implies that $\mathbf{y}\in {\cal
F}_{\frac{\gamma}{\lambda}+1}$. Hence, by Lemma \ref{lem1},
$w_P(\mathbf{x})=w_P(\mathbf{y})\ge \frac{\gamma}{\lambda}+1$, and
equality holds if and only if $\mathbf{y}$ has exactly
$\frac{\gamma}{\lambda}+1$ non-zero components with value
${\lambda} /(\gamma+\lambda)$. In that case,
$(\frac{\gamma}{\lambda}+1)\mathbf{y}=(\frac{\gamma}{\lambda}+1)\mathbf{x}/\sum_{j=1}^n
x_j$ must be a binary vector, say $\mathbf{c}$, with weight
$\frac{\gamma}{\lambda}+1$.

Now, we show that $\mathbf{c}$ must be a codeword of $C$. Since
$\mathbf{c}\in K(H)$, ${\rm{supp}}(\mathbf{c})$ is a stopping set of
$H$, i.e., the restriction of $H$ to ${\rm{supp}}(\mathbf{c})$, say
$H(\mathbf{c})$, has no rows of weight one. Note that any two
distinct columns of $H(\mathbf{c})$ have at most $\lambda$ common
$1$'s. Suppose $\mathbf{b}$ is a non-zero row of $H(\mathbf{c})$ and
the $j$-th component of $\mathbf{b}$ is $1$, where
$j\in\{1,2,\ldots,\frac{\gamma}{\lambda}+1\}$. Since the $j$-th
column of $H(\mathbf{c})$ has at least $\gamma$ $1$'s, there exists
a $\gamma\times (\frac{\gamma}{\lambda}+1)$ matrix, say
$H(\mathbf{c},j)$, consisting of $\mathbf{b}$ and other $\gamma-1$
rows of $H(\mathbf{c})$ such that the $j$-th column of
$H(\mathbf{c},j)$ is the all-$1$ column. Since any two distinct
columns of $H(\mathbf{c},j)$ have at most $\lambda$ common $1$'s,
each of the columns other than the $j$-th column has at most
$\lambda$ $1$'s. Now we count the number of $1$'s in
$H(\mathbf{c},j)$, say $\Delta$, in two ways. From the view of
columns, $\Delta\le
{\gamma}+{\lambda}\frac{\gamma}{\lambda}=2\gamma$. From the view of
rows, since ${\rm{supp}}(\mathbf{c})$ is a stopping set, each row of
$H(\mathbf{c},j)$ has at least two $1$'s, which implies $\Delta\ge
2\gamma$. Thus, $\Delta= 2\gamma$, and every row of
$H(\mathbf{c},j)$ has exactly two $1$'s, which implies
$w_H(\mathbf{b})=2$. Hence, the weights of rows of $H(\mathbf{c})$
are either $0$ or $2$, which implies that $\mathbf{c}$ satisfies
every parity-check equation in $H$ and thus is a codeword.

\section{Conclusions}

In this correspondence, we study the minimum pseudo-weight and
minimum pseudo-codewords of LDPC codes. We characterize the
pseudo-codewords of an LDPC code which attain the lower bound $d_L$
of Kelley, Sridhara, Xu, and Rosenthal on the minimum pseudo-weight.
That is, the pseudo-weight of a pseudo-codeword of an LDPC code is
equal to $d_L$ if and only if this pseudo-codeword is a real
multiple of a codeword with weight $d_L$. Furthermore, it is shown
that if the minimum distance of this LDPC code is equal to $d_L$,
then the minimum codewords, the nonempty stopping sets of smallest
size and the minimum edges are all equivalent, which implies that LP
decoding is asymptotically optimal for this LDPC code. Then, we show
that the lower bound of Kashyap and Vardy on the stopping distance
of an LDPC code is also a lower bound on the pseudo-weight of a
non-zero pseudo-codeword of an LDPC code with girth 4. The same
characterization results mentioned above for the lower bound of
Kelley, Sridhara, Xu, and Rosenthal are also obtained for this new
lower bound on the minimum pseudo-weight. Some LDPC codes are listed
to illustrate these results. Finally, we pose a further research
problem: For a binary LDPC code $C$, construct a parity-check matrix
$H$ with minimum number of rows such that the minimum pseudo-weight
of $C$ is equal to the minimum distance of $C$, and the number of
minimum edges is equal to the number of minimum codewords of $C$,
i.e., LP decoding is asymptotically optimal for this LDPC code.
Until now, we do not even know whether such a parity-check matrix
exists for every binary linear code.

\section*{Acknowledgment}
The authors would like to thank Dr. Sridhara for kindly affording
the preprint \cite{ks}, and Mr. X. Ge for his calculation of the
minimum distance in Example 6 by computer. The authors wish to
express their appreciation to the three anonymous reviewers and
Associate Editor Marc Fossorier for their valuable suggestions and
comments that helped to greatly improve the correspondence.

\end{document}